\documentclass[ reprint,twocolumn,10pt,superscriptaddress, aps,prmaterials]{revtex4-2}

\usepackage{amsmath,amssymb,amsfonts,csquotes}
\usepackage{graphicx}
\usepackage{xcolor}
\usepackage{float}

\usepackage{hyperref}
\hypersetup{
colorlinks=true,
urlcolor= blue,
citecolor=blue,
linkcolor= blue,}

\usepackage{bm}

\begin{document}

\title{Observation of banded spherulite in a pure compound by rhythmic
 growth}
\author{Subhadip Ghosh}
\author{Dipak Patra}
\author{Arun Roy}

\email[]{aroy@rri.res.in}
\affiliation{Soft-Condensed Matter Group, Raman Research Institute, 
Bangalore 560080, India}
\date{\today}

\begin{abstract}
Banded spherulitic growth of crystal is observed in some materials with 
spherically symmetric growth 
front and periodic radial variation of birefringence. This variation of 
birefringence in quasi two dimensional geometry produces 
concentric interference colour bands when viewed through crossed polarisers. 
In most materials, the banded spherulites are found to be formed by 
radially oriented periodically twisted fibrillar 
crystallites. Here, we report the formation of  banded spherulites due to 
the rhythmic growth of concentric crystallite-rich and 
crystallite-poor bands for a pure compound consisting of strongly polar rod 
like molecules.
The compound exhibits coexistence of untwisted fibrillar crystallites and 
an amorphous phase in its most stable solid 
state. On sufficient supercooling of the sample from its melting point, the 
banded spherulites are formed with a periodic 
variation of composition of untwisted radially aligned fibrillar 
crystallites and an amorphous solid phase. We have developed 
a time dependent Ginzburg-Landau model to account for the observed banded 
spherulitic growth.
\end{abstract}

\maketitle

\section{Introduction}
The spherulitic growth of solid is a 
ubiquitous phenomenon exhibited by many 
different types of materials
such as polymers \cite{crist2016,lotz2005}, minerals 
\cite{lofgren1974,coish1979,kirkpatrick1974}, elements
\cite{minkoff1966,bisault1991}, metals \cite{shtukenberg2012} 
and salts \cite{beck2010,oaki2004,thomas2012}. 
It is also found in biological materials
like coral skeletons \citep{sun2017}, kidney stones \citep{atar2010}, 
proteins \cite{chow2002,coleman1960,krebs2004} 
and urinary sediments \citep{catalina1970}.   
In spherulitic growth, the solid phase after nucleation grows with 
a spherical growth front with continuous orientational symmetry in contrast to the growth of a crystal having
discrete orientational symmetries. The spherulites were initially named 
as “circular crystals” for its circular boundary 
in quasi two dimensional geometry which changed later to 
spherulite \citep{brewster1853}.  
In spite of research on it over a century, the detail 
understanding of the mechanism of this abundantly found natural growth phenomena is still 
incomplete \cite{shtukenberg2012}. 
The spherulitic  growth morphology is often associated with the 
formation of many radially aligned fibrillar crystallites which branch 
non-crystallographically for filling space during the growth 
\cite{keith1963}. This 
distinct characteristic of spherulites separates it from the other 
polycrystalline aggregates. 

For some materials, the spherulitic growth is accompanied by a series of 
equidistant concentric bands and is 
known as banded spherulite. In quasi two dimensional geometry, the banded 
spherulite domain appears as a 
flattened disk with circular boundary 
and these bands appear as concentric 
circular rings. The polarizing optical 
microscopic textures of these spherulites show concentric bands with 
periodic variation of
interference colour. Polymers generally form 
this kind of banded spherulite
\cite{shtukenberg2012, keith1984, crist2016, lotz2005}. 
Some small molecular systems also produce banded spherulites
\cite{shtukenberg2010hip, shtukenberg2012man, pisula2004dlc, 
cui2013aspirin, shtukenberg2011testo,
lin2006tlc, bechhoefer1997, bechhoefer2000}.  
An organized periodic twisting of radially 
aligned fibrillar crystallites in the banded spherulitic domain has been 
observed to generate a periodic change of effective 
birefringence along the radial direction. This undulation in the 
birefringence gives rise to concentric bands with 
periodic variation of colours when viewed through 
crossed polarisers. In contrast, some 
multicomponent blends 
of materials form banded 
spherulites by rhythmic deposition of host and additive 
materials \cite{wang1997, chen2005, chen2007}.
Even in some of these cases, twist order of fibrils  
has been found to contribute to the modulation of effective birefringence 
along the radial direction.
So the organised twisting of the fibrillar 
crystallites is considered to be 
a primary mechanism for the  formation
of the banded spherulites.

In this work, we report the formation 
of banded spherulite by a pure liquid 
crystalline compound consisting of 
relatively small rod-like molecules.  
We show that the  rhythmic growth of 
concentric crystallite-rich and 
crystallite-poor amorphous zones is the underlying mechanism for the 
formation of banded spherulites of this compound contrary to the organised twisting of the fibrillar 
crystallites.
The crystallite-rich zones have higher density of radially aligned 
fibrillar nano crystallites while 
crystallite-poor zones are rich in solid amorphous state of the compound.
The alternation of these concentric regions changes the effective 
birefringence periodically along the 
radial direction of the spherulite giving rise to the concentric 
interference colour bands between crossed polarisers.
We also develop a phase field model 
to account for the formation of banded spherulite due to rhythmic growth in our system.

\section{Experiment}
The commercially available liquid crystalline compound 
4\'{}-octyloxy-4-cyanobiphenyl (8OCB) was used with 99$\%$
purity. 
For spherulitic growth, the sample sandwiched between two cleaned 
coverslips in its isotropic 
phase was cooled to a desired temperature below its melting point using a 
microscope hot stage (Linkam). The banded 
spherulitic domains of the sample were generally formed above the 
supercooling $\Delta$T = 22.5 K. 
The spherulitic domains grown in this way were used for polarising optical 
microscope (POM) observations 
and Raman spectroscopic studies. 
The spacing between the spherulitic bands was determined by taking 
photographs using a digital camera
(Canon EOS 80D) attached to the polarising microscope (Olympus BX50) 
followed by analysis of the images using the 
imageJ software. 

The micro Raman spectroscopic studies of the spherulite domain were 
performed using a Raman 
spectrometer (Jabin-Yvon T64000) 
equipped with a nano-positioning stage. An 
air cooled Argon 
laser (Melles Griot) beam of  wavelength 
514 nm  was used as an excitation source. 
An objective lens with 50X magnification and 0.75 numerical aperture was 
used to focus the beam on the sample. 
The Raman spectra of the sample were recorded from the backscattered light.
The theoretically estimated diameter 
of the focused spot is less than 1 $\mu$m. 

The field emission scanning electron microscopy (FESEM) studies of the 
spherulite domain were conducted using the 
CARL ZEISS (ULTRA PLUS model) system. The banded spherulites of the sample 
were formed from its melt on 
cleaned ITO coated glass plates at room temperature. The free surface of 
these samples
was sputter coated with platinum [QUORUM (Q150R S)] for FESEM imaging. For 
the cross sectional view of the domain, 
the spherulites formed on coverslips were cleaved in the middle and then 
the cross sectional areas were sputter coated 
with platinum for FESEM imaging.

The fluorescent dye (Rhodamine 6G)  was mixed with the 8OCB sample at 
1:10$^4$ wt/wt ratio. 
The mixture was stirred for 3 hours above the  clearing temperature of 8OCB 
for homogeneous mixing. 
The sample was then sandwiched between two coverslips and supercooled to 
room temperature to form the 
banded spherulites. The fluorescent images of the spherulite domain were 
captured by a confocal 
microscope (Leica SP8) equipped with an Argon gas laser. The exciting 
wavelength was 514 nm.

A DY 1042-Empyrean (PANalytical) X-ray diffractometer with PIXcel 3D 
detector was used to acquire the 
X-ray diffraction (XRD) profiles of the sample using CuK$\alpha$ radiation 
of wavelength 1.54 \text{\normalfont\AA}.
The banded spherulites were formed on a cleaned coverslip on cooling the 
sample to room temperature from
its isotropic phase and its formation was confirmed by POM observation. The 
sample was then scraped off this substrate 
and placed on a silicon plate.  The XRD measurements were performed in the 
grazing angle of incidence of the X-ray 
beam. The silicon plate has flat XRD profile which does not interfere with 
that from the sample.

The temperature variation of dielectric constant of the sample was measured 
using a homemade setup. The detail of 
this setup has been discussed elsewhere \cite{ghosh2021}.  A commercially 
available liquid crystal cell (INSTEC Inc.) 
of sample thickness 5 $\mu$m 
and electrode area 5$\times$5 mm$^2$ was used for dielectric measurement. 
Prior to filling the sample, the 
capacitance of the empty cell was measured. The cell was then filled with 
the sample in its isotropic phase by
capillary action on a hot stage. After that, it was supercooled to the 
lower temperature to form banded spherulite domains.
The formation of the banded spherulites was confirmed using POM observation 
of the sample.
A sinusoidal voltage of rms amplitude 0.5 V and frequency 5641 Hz was used 
for the dielectric measurements.
The ratio of the capacitance of the filled cell to the empty cell gives the 
dielectric constant of the sample.

\section{Results and discussions}

The liquid crystalline compound 4\'{}-octyloxy-4-cyanobiphenyl 
abbreviated as 8OCB exhibits a variety of 
crystal and liquid crystal phases. The sample shows a melting transition to 
the smectic~A phase at 327.6 K
on heating from its most stable crystal phase known as commercial powder or 
CP phase.  On further heating, the 
compound transforms to nematic and to isotropic liquid phase at 340.1 K and 
353.1 K respectively.  
The sample exhibits a large range of supercooling from 
its smectic~A phase. 
The banded spherulitic domains are formed on supercooling the smectic phase 
by at least 22.5 K below the melting temperature. The probability of 
getting the banded spherulites 
increases with higher supercooling. 

These banded spherulites generally show four 
black brushes parallel to the polarisers forming a Maltese cross in 
addition to the concentric interference 
colour bands between crossed polarisers. These colour bands arise from the periodic variation of the 
effective birefringence in the sample along 
the radial direction. Fig.~\ref{fig1}a shows a polarising optical 
microscope (POM) texture of a banded 
spherulite formed  between two coverslips on cooling the sample to room 
temperature.  The black and white appearance 
of the interference colour bands in this texture is due to relatively lower thickness of the 
sample with phase retardation 
in the first order region of the Levy chart. The dark arms of the Maltese 
cross remain invariant on rotating the sample between crossed polarisers.
This implies that the major axis of the effective refractive index ellipse in 
the plane perpendicular to the light path lies 
parallel or perpendicular to the radial direction of the spherulitic 
domain.

The POM studies using a $\lambda$-plate (530 nm) were performed to 
determine the orientation of this major
axis. Fig.~\ref{fig1}b is the POM image after introducing the $\lambda$-
plate in the optical path of the microscope. 
The yellow double headed arrow shows the orientation of the slow axis of 
the $\lambda$-plate 
with respect to the orientations of the crossed polarisers. It can be 
inferred from the Levy chart that there is effective addition 
and subtraction of path differences due to the sample and 
the $\lambda$-plate in the blue and yellow coloured regions 
of the texture respectively. This observation clearly indicates that the 
major axis is oriented parallel to the azimuthal 
direction of the banded 
spherulite. Hence the spherulite observed for this sample is optically 
negative \cite{shtukenberg2012, crist2016}.
The white double headed arrows in Fig.~\ref{fig1}b show the orientation of 
the major axis
around the seed of this spherulite domain.  The spacing between successive 
bands is also measured for different 
values of supercooling of the sample and compared with the theoretical 
results discussed later. The band spacing 
does not vary appreciably with the thickness
of the sample (see Fig.~S1 
in the Supplementary Material) 
which also agrees with our theoretical
model.

\begin{widetext}
\begin{minipage}{\linewidth}
\begin{figure}[H]    
2    \includegraphics[width=15cm]{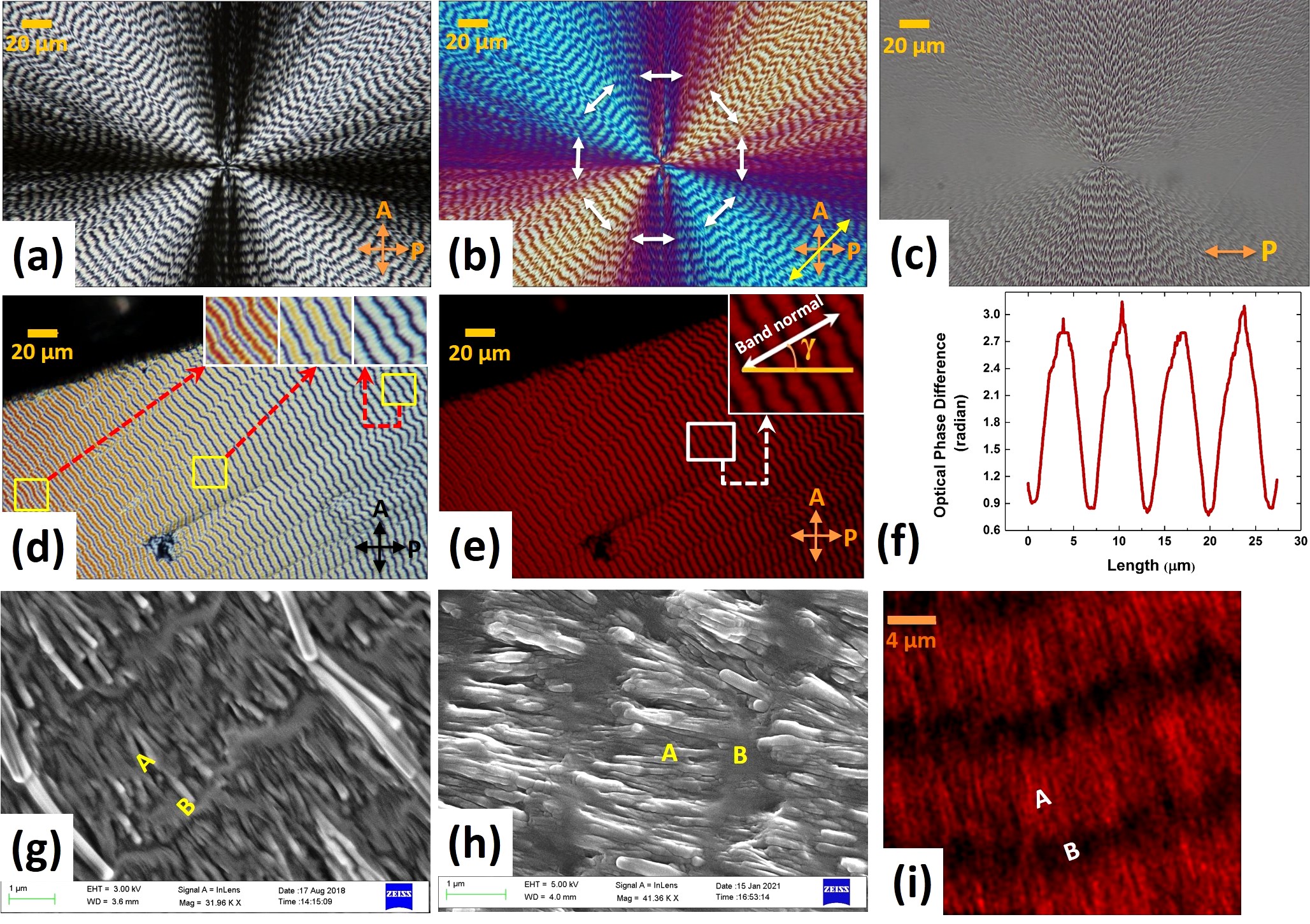}
    \centering
    \caption{\textbf{Microscopic textures of banded spherulite.} 
    The POM textures of banded spherulite of compound 8OCB  at 
    room temperature (a) between crossed 
    polarisers, (b) between crossed polarisers and a $\lambda$-plate and 
    (c) with only the polariser in 
    the optical path. The white double headed arrows in (b) indicate the 
    orientations of major refractive index around the 
    seed. (d) The POM texture of a banded spherulite domain between crossed 
    polarisers for relatively thicker sample 
    at room temperature. The insets show the variation of colour in the 
    bands in three different regions with slightly 
    different sample thickness. (e) The same as in (d) with a red filter 
    introduced in the optical path. The inset shows the 
    variation of light intensity along the band normal which makes an angle 
    $\gamma$ with the polariser. (f) The variation of 
    optical phase difference along the radial direction of the banded 
    spherulite shown in (e). The FESEM image of (g) the 
    top surface of a banded spherulite and (h) the cross-sectional view of 
    a banded spherulitic obtained from the 
    cleaved surface showing the bulk structure. (i) The fluorescent image 
    of a dye doped banded spherulite at room 
    temperature.  The regions marked \enquote{A} and \enquote{B} represent 
    the concentric crystallite-rich and 
    crystallite-poor bands respectively.}
    \label{fig1}
\end{figure}
\end{minipage}
\end{widetext}

Fig.~\ref{fig1}c is the bright field image of the sample shown in 
Fig.~\ref{fig1}a with only the polariser in the light
path. In this figure, the bands have low (high) optical contrast where band 
normal is parallel (perpendicular) to the polariser.
These observations indicate that the radial component of refractive index 
does not vary appreciably 
along the radial
direction but the strong variation of the azimuthal component along the 
radial direction produces the undulation of
the birefringence in the spherulite. 

For relatively thicker sample, there are multiple interference colour rings 
in a given spherulitic band as 
shown in Fig.~\ref{fig1}d. The insets in Fig.~\ref{fig1}d are the magnified 
view of these 
interference colour rings in three regions with slightly different sample 
thickness. 
The variation of phase retardation in the texture can be measured from the 
intensity variation by
introducing a red filter in the light path of the POM. 
Fig.~\ref{fig1}e is the POM image of the spherulite domain with a periodic 
variation of
intensity along the radial direction of the spherulite. 
The inset in Fig.~\ref{fig1}e depicts the magnified view of the marked 
region of this
domain where the band normal makes an angle $\gamma$ = 30.5$^{\circ}$ with 
respect to the 
polariser. Then the phase retardation can be calculated from the
formula $I=I_0\sin^2 (2\gamma)\sin^2 (\phi /2)+I_d$,
where $I_0$, $I$ and $I_d$ are the incident, measured and the dark 
background intensities
with crossed polarisers respectively. The phase retardation $\phi=2\pi 
d\Delta n/\lambda$, where $d$ is the 
sample thickness, $\Delta n$ is the effective linear birefringence of the 
sample and $\lambda$ is the wavelength of incident light.
Fig.~\ref{fig1}f shows the measured variation of optical phase retardation 
along the radial direction of the spherulite.
The variation of sample thickness for the selected small region shown in 
the inset of Fig.~\ref{fig1}e is negligible and
the oscillation of phase retardation across the bands mainly arises due to 
the variation of the birefringence in the sample.

The field emission scanning electron microscopy (FESEM) studies of the 
banded spherulite domain showed 
concentric bands 
with same periodicity as found in POM studies. The high resolution FESEM 
texture of the spherulitic domain 
revealed that it consists of numerous fibrillar nano crystallites embedded 
in an amorphous phase. The bands in the
spherulite arise from the periodic growth of crystallite-rich and 
crystallite-poor concentric zones as shown in 
Fig.~\ref{fig1}g. Similar repetitive structure 
was also found in the FESEM texture of the bisected
cross sectional surface area of the sample 
(Fig.~\ref{fig1}h) showing its bulk nature 
(also see Fig.~S2 in the Supplementary Material). 
The FESEM textures of the sample revealed that the 
crystallite-rich zones (marked as 
\textbf{A}) have high density of 
fibrillar crystallites aligned along the radial direction 
while crystallite-poor zones (marked as 
\textbf{B}) are rich in solid 
amorphous state of the compound.
The crystallites have length similar to the periodicity of the bands while 
the width varies from 100 nm to 300 nm.

Some polymers and a few small molecular systems exhibit concentric colour 
banded spherulites \cite{shtukenberg2012,
lotz2005, keith1984, crist2016, shtukenberg2010hip, pisula2004dlc, 
shtukenberg2012man, cui2013aspirin, 
shtukenberg2011testo} in which organised 
twisting of fibrillar units along their 
long axis has been 
observed as the mechanism of the band formation. In some 
small molecular systems, 
a periodic variation of orientation of individual crystallites has been 
reported to produce concentric colour bands
\cite{shtukenberg2011, woo2016}.  But in our sample, neither twist nor 
periodic change of orientation of fibrillar 
crystallites is found. The FESEM studies depict that the rhythmic growth of 
concentric crystallite-rich  and crystallite-poor 
zones produces the banded spherulite of 8OCB.
The periodic change in birefringence along the radial direction due to 
these concentric zones gives rise to 
the interference colour bands observed between crossed polarisers. The 
crystallite-poor regions possess least 
birefringence due to their predominant amorphous nature while the 
crystallite-rich zones have larger birefringence due to the presence of 
large number of radially aligned crystallites.

\begin{widetext}
\begin{minipage}{\linewidth}
\begin{figure}[H]
   \includegraphics[width=14cm]{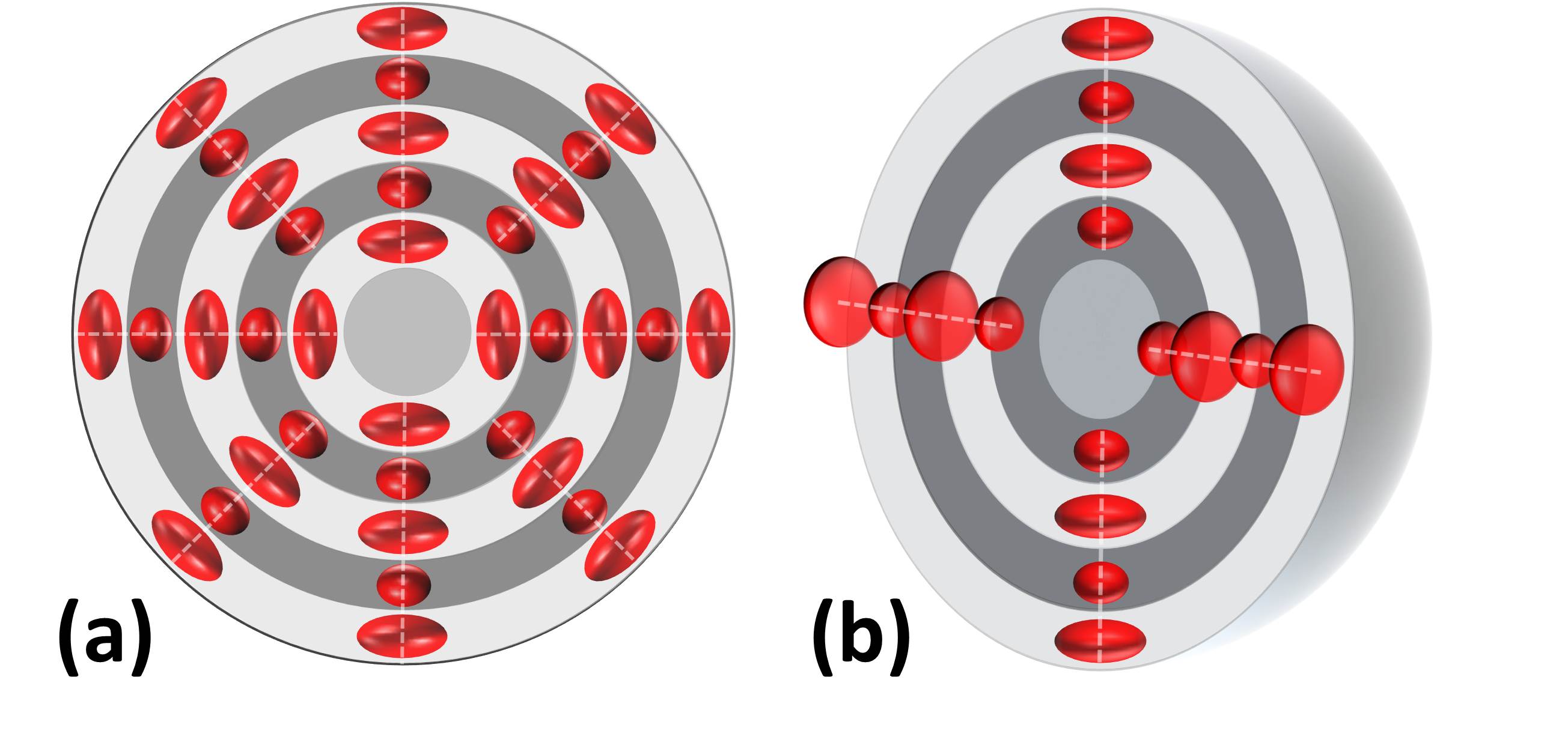}
   \centering
    \caption{\textbf{Indicatrix ellipsoid in banded spherulite.}
    The schematic representation of the variation of optical 
    indicatrix along the radial direction of the banded 
   spherulite of the compound 8OCB. The (a) top  and (b) side views showing 
   the uniaxial oblate indicatrix with radially 
   oriented optic axes.} 
    \label{fig2}
\end{figure}
\end{minipage}
\end{widetext}

A fluorescent image of the banded spherulite domain (see Fig.~\ref{fig1}i) 
was obtained by doping the 
sample with a fluorescent dye (Rodamine 6G). It shows a periodic variation 
of dye concentration 
along the radial  direction of the spherulite domain. The crystallite-rich 
zones are found to have higher concentration of 
dye molecules compared to the crystallite-poor zones which makes these 
zones brighter and darker respectively  
in the fluorescent image. The dynamic advection of the dye molecules 
during the growth of crystallite-rich 
and crystallite-poor zones perhaps can explain their distribution in the 
banded spherulite of the sample.
The molecular density is higher in crystallite-rich zones compared to 
crystallite-poor amorphous regions. 
Thus the growth of a crystallite-rich band produces a diffusive current 
which advects the dye molecules from the leading 
edge of the growth front  that gives rise to the higher density of the dye 
molecules in this band.  The depletion of 
molecular density promotes the formation of crystallite-poor amorphous 
zones with lower density of the dye molecules. 
The repetition of these processes leads to the formation of banded 
spherulite in this sample.

Fig.~\ref{fig2} shows the schematic representation of the variation in the 
refractive index ellipsoid along
the radial direction of the banded spherulite. The FESEM studies revealed 
that the elliptical or rectangular cross sections 
of radially aligned fibrillar crystallites are oriented randomly in the 
banded spherulite giving rise to a
uniaxial structure about the radial direction (see  Fig.~S3 in the Supplementary Material). The POM studies showed that
the azimuthal component of the refractive index is larger than the radial 
component in the spherulite. Hence, the 
banded spherulite of 8OCB possesses uniaxial oblate indicatrix with a 
radially oriented optic axis as 
shown in Fig.~\ref{fig2}. 
The radial component of refractive index does not vary appreciably along 
the radial direction but the periodic variation
of the transverse component gives rise to the modulation in the 
birefringence of the sample as observed in 
Fig.~\ref{fig1}c.

The X-ray diffraction (XRD) studies were performed to identify the crystal 
structure of the banded spherulite domain. The XRD profile 
(Fig.~\ref{fig3}a) of banded spherulite is identical to 
that of the commercial powder
or CP crystal phase of 8OCB. We have recently shown that the CP crystal 
phase is composed of fibrillar 
nano crystallites of the compound 
embedded in its own amorphous solid state \cite{ghosh2021}.
All the peaks 
observed in the XRD profile are associated with the monoclinic crystal 
structure of the nano crystallites. The 
amorphous component shows only 

\begin{widetext}
\begin{minipage}{\linewidth}
\begin{figure}[H]
    \includegraphics[width=18cm]{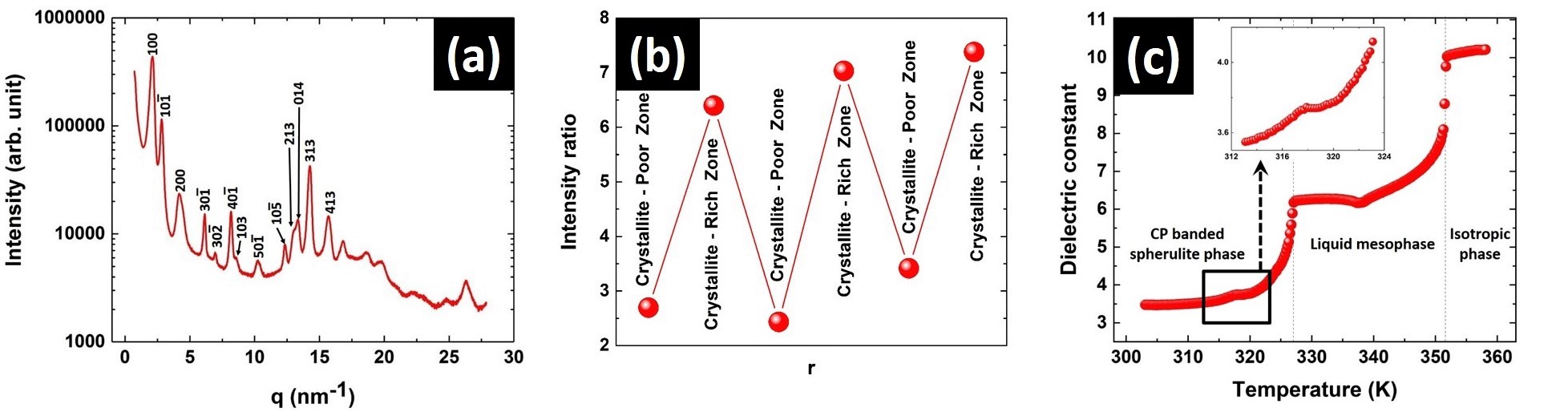}
    \centering
    \caption{\textbf{Two phase coexistence from XRD, 
    Raman and dielectric profile.}
    (a) The XRD profile of the banded spherulite of the compound 
    8OCB.
    (b) The variation of intensity ratio of the Raman peaks associated with 
    the CN group of the molecules at 
    crystallite-rich and crystallite-poor zones of the banded spherulite 
    along its radial direction.
    (c) The variation of dielectric constant of the banded spherulite with 
    temperature during heating of the sample. 
    The inset highlights the step change in dielectric constant  at 318.1
     K.}
    \label{fig3}
\end{figure}
\end{minipage}
\end{widetext}
a broad hump from 11 nm$^{-1}$ to 21 
nm$^{-1}$ in this XRD profile. 
Thus the XRD studies indicate that the banded spherulites of the compound 
has similar composition as its CP crystal 
phase. 

The spatially resolved variation in the compositions of the sample in the 
banded spherulitic domain 
was probed by micro Raman spectroscopy. The stretching vibration peak at 
2235 cm$^{-1}$ of the nitrile (CN) bond 
of 8OCB molecules was used to examine the compositions of the spherulite 
along its band normal direction. 
This Raman peak 
is reported to be quite sensitive to the molecular environment around the 
CN bond of this molecule \cite{hori2000}. It
is asymmetric in the spherulite domain and can be fitted 
with two closely spaced peaks (see Fig.~S4 in the Supplementary Material). 
The fitted intense peak positioned 
at 2235 cm$^{-1}$ arises from  the nano crystallites of the banded 
spherulite and less intense 
shoulder peak positioned at 2226 cm$^{-1}$ originates from the amorphous 
component \cite{ghosh2021}. The ratio of the
intensities of these peaks obtained in the crystallite-rich and 
crystallite-poor zones along the radial direction
of the banded spherulite is shown in Fig.~\ref{fig3}b. The higher and lower 
values of this intensity ratio
at crystallite-rich and crystal-poor zones respectively indicates the 
periodic 
variation of the compositions along the radial direction of the banded 
spherulite.

The temperature variation of dielectric constant of the banded spherulite 
sample during heating it 
from room temperature 
is shown in Fig.~\ref{fig3}c. A sudden change of slope in the dielectric 
profile (see inset in Fig.~\ref{fig3}c) 
was observed at 318.1 K indicating a transformation in the sample. But no 
change in the POM textures 
of the banded spherulites was found at this temperature. We attribute it to 
the softening of the amorphous
component of the sample.  On heating, the bands started to become blur 
irreversibly at 321.1 K and the texture 
transformed completely to non-banded spherulite at 325.1 K. 
The FESEM studies of this non-banded spherulite revealed that the lengths 
of the radially aligned fibrillar 
crystallites were increased and their distribution was uniform along the 
radial direction giving rise to the non-banded 
texture (see  Fig.~S5 in the Supplementary Material ). 
Thus the amorphous component after softening 
gradually transforms to the crystallites and smectic phase on heating which 
accounts for the observed increase in
dielectric constant with increasing temperature before complete melting of 
the sample to the smectic phase at 
327.6 K \cite{ghosh2021}.

\subsection*{Theoretical Model}
In the last few decades, a few theoretical models based on coherent twisting of fibrils have 
been introduced to account for banded spherulite. The  twisting can occur in a 
system due to multiple  causes such as surface
stress mismatch \cite{keith1984, lotz2005}, isochiral screw dislocation 
\cite{patel2002, toda2001}, auto deformation 
\cite{shtukenberg2012}, 
self-induced concentration or mechanical fields on growth 
kinetics \cite{schultz2003} and topological 
defects \cite{hatwalne2010}.  In addition some phase field models have also been developed to understand 
this type of spherulitic growth \cite{granasy2005,fang2015}. 
However, coherent twisting of the fibrillar crystallites has
not been observed in the banded spherulite for this pure liquid crystal compound.

Armed with our experimental results, a time dependent Ginzburg-Landau (TDGL) 
model C is developed to account for the 
rhythmic growth of banded spherulitic domain observed in our system. 
This type of model has 
been successfully applied 
in various physical growth phenomena in different systems 
\cite{wheeler1992, elder1994}. 
Kyu \textit{et al.} simulated ring banded structure 
using this model and emphasized rhythmic growth assisted banded spherulite
formation in polymer blends \cite{kyu1999,kyu2006}.
The model describes the dynamics of a system using a conserved 
and a non-conserved order parameters. 
Our experimental studies have clearly established that the growth of 
alternating concentric
crystallite-rich and crystallite-poor amorphous zones gives rise to the 
banded spherulites of 8OCB. Therefore, 
we define the conserved order parameter $\phi=[\rho-\rho_0]/\rho_0$ which 
describes the local 
deviation of density in these zones from the average density $\rho_0$ of 
the smectic phase.
In addition, the non-conserved
order parameter $\psi=(\rho_{nc}-\rho_{a})/\rho_0$ describes the local 
composition, where $\rho_{nc}/\rho_0$ and 
$\rho_{a}/\rho_0$ are the fractional densities of molecules in the nano 
crystalline and
amorphous solid phase respectively. The smectic phase with $\psi=0$ 
undergoes a first order transition to
the spherulitic domain with $\psi\neq 0$.

Using these order parameters, the free energy density of the system in its 
smectic phase can be written as
\begin{multline}
f = \frac{A}{2}\phi^2 +\frac{1}{2}k_\phi |\nabla \phi|^2 
+ W \Big[ \frac{\alpha (T)}{2}\psi^2 - \frac{1+\alpha(T)}{3}\psi^3 +
\frac{1}{4}\psi^4 \Big]\\
+\frac{1}{2}k_\psi |\nabla \psi|^2-\gamma \phi \psi
\label{eq.fed}
\end{multline}
The first two terms in equation~\ref{eq.fed} represent the free energy
associated with the order parameter $\phi$ of the sample. We assume that $
\phi$ is non-critical across the transition 
to the spherulitic state and only the terms upto the quadratic order are 
retained. The third and fourth terms 
represent the free energy associated with the order parameter $\psi$
describing the first order phase transition from the smectic phase to the 
spherulitic state. 
Here, $0\leq\alpha(T)= (T-T_c)/(T_m-T_c)\leq 1$ is the temperature 
dependent parameter
driving this first order transition and $T_m$, $T_c$ are the melting 
temperature and 
supercooling limit of the smectic phase respectively.  
The last term in equation~\ref{eq.fed} is the lowest order coupling between 
these order parameters. 

Using this free energy expansion, the dimensionless form of the TDGL 
equations for the system can 
be written as
\begin{eqnarray}\label{TDGLphieqn}
\frac{\partial \phi}{\partial t} &=& \nabla^2 \Big[\phi -\frac{\gamma}{A}
\psi -\epsilon \nabla^2\phi  \Big]\\
\epsilon\frac{\partial \psi}{\partial t} &=& -\frac{\xi^2_\phi}
{\lambda^2_D} \Big[\psi(\psi-1)(\psi-\alpha)
-\frac{\gamma}{W}\phi -\epsilon \frac{\xi^2_\psi}{\xi^2_\phi}\nabla^2\psi  
\Big]\label{TDGLpsieqn}
\end{eqnarray}
where $\epsilon=(\frac{\gamma^2}{WA} -\alpha)$ is a dimensionless parameter and
$\xi_\phi=\sqrt{\frac{k_\phi}{A}}$, 
$\xi_\psi=\sqrt{\frac{k_\psi}{W}}$ and $\lambda_D = \sqrt{\frac{A\Gamma_\phi}{\Gamma_\psi W}}$ have the dimension of length. Here $\Gamma_\phi$ 
and  $\Gamma_\psi$ are two dynamical coefficients of the system.  The 
detail derivation of 
these equations is given in the Supplementary Material .

The linear stability analysis (LSA) of equation~\ref{TDGLphieqn} and 
equation~\ref{TDGLpsieqn} was 
performed to investigate the growth of  banded spherulite on 
quenching the system from higher temperature smectic phase. The TDGL 
equations were linearised
with respect to small perturbations  $\phi(\vec{r},t)= \delta\phi_0 
e^{\sigma t} e^{i\vec{q}\cdot \vec{r}}$, 
$\psi(\vec{r},t)= \delta\psi_0 e^{\sigma t} e^{i\vec{q}\cdot \vec{r}}$  
about the smectic phase ($\phi=0,\psi=0$).
The resulting eigenvalue equations (see  Supplementary Material ) 
determine the stability of the 
smectic phase with respect to these perturbative modes for different values 
of the control parameter $\alpha$.
The modes with negative real part of $\sigma$ 
decay to zero and are therefore stable while the modes with positive real 
part of $\sigma$ grow with time and are unstable. 

Fig.~\ref{fig:Sigma}a shows the variation of $\sigma$ with the wave vector 
$q$ for different values of $\alpha$. 
As can be seen from Fig.~\ref{fig:Sigma}a, $\sigma$ is positive for modes 
with wave vector $q$ lying between zero 
and $q_{max}$ and these unstable modes give rise to the formation of banded 
spherulite in this system.
Among these unstable modes, $\sigma$ has the highest positive value for 
$q=q_c$ and this most
unstable mode will be dominant during the growth of the banded spherulite.
Fig.~\ref{fig:Sigma}b shows the variation of $q_{max}$ and $q_c$ with the 
parameter $\alpha$ obtained from LSA.
Though $q_c$ varies strongly with $\alpha$, the $q_{max}$ does not vary 
appreciably with it.
The wavelength corresponding to this dominant mode calculated from the LSA 
is compared with
the experimentally measured band spacing of the banded spherulite as shown 
in Fig.~\ref{fig:Sigma}c for
different values of supercooling. The theoretical result agrees well with 
the experimental data confirming the general 
validity of this model.

\begin{widetext}
\begin{minipage}{\linewidth}
\begin{figure}[H]
   \includegraphics[width=17.3cm]{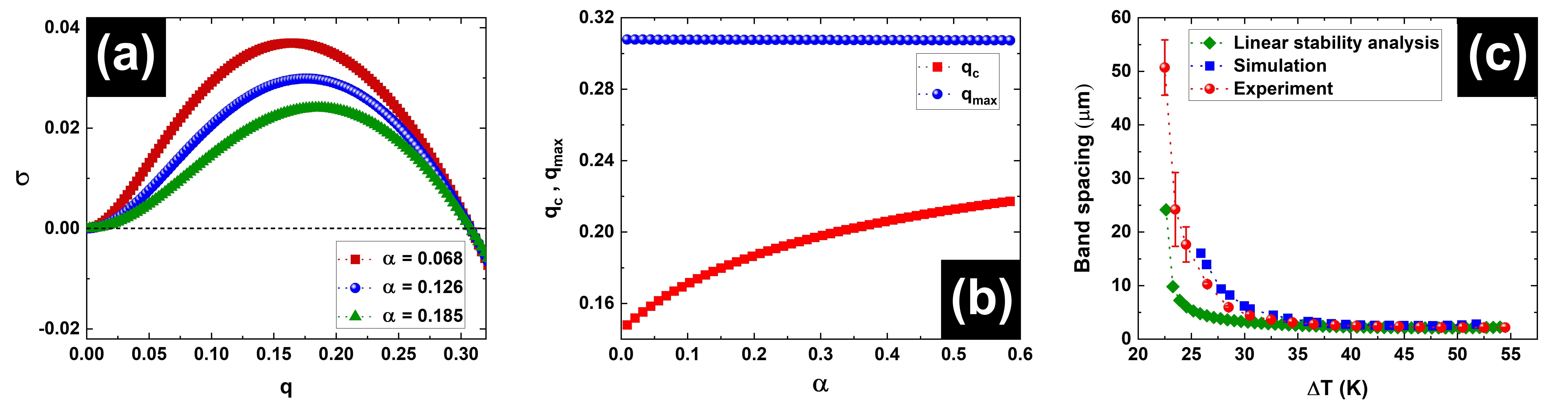}
   \caption{\textbf{Effect of supercooling on banded spherulite.} 
   (a) The variation of $\sigma$ with the  wave vector $q$ for 
   different values of $\alpha$. The modes in the 
   wave vector range $0<q<q_{max}$ are unstable. (b) The variation of 
   $q_{max}$ and $q_c$ as a function of $\alpha$. 
   (c) Comparison between experimental, linear stability analysis and 
   simulation results for the variation of band 
   spacing with supercooling $\Delta$T. The values of the parameters used 
   for the theoretical model are 
   $\gamma /A = 0.242$, $W/A=0.1$, $\xi_\phi=0.04$ $\mu$m, $
   \xi_\psi=3.16\xi_\phi$, $\lambda_D=1.41\xi_\phi $.}
   \label{fig:Sigma}
\end{figure}
\end{minipage}
\end{widetext}

A general analytical method of finding solutions of the coupled nonlinear 
TDGL equations given by 
equations~\ref{TDGLphieqn} and \ref{TDGLpsieqn} is not known. These 
equations were solved numerically under 
no flux boundary condition using a finite difference method in two 
dimension.
The numerical solutions of the order parameters $\phi$ and $\psi$ for a 
banded spherulite 
growing from a seed at
the centre are shown in Fig.~\ref{fig:FDSimEg0}a and 
Fig.~\ref{fig:FDSimEg0}b respectively (also see Fig.~S6 in the Supplementary Material).
The formation of the
ring banded structure is clearly observed as found experimentally. 
Fig.~\ref{fig:FDSimEg0}c shows the graphical
profiles of the order parameters along a radial direction of the 
spherulite. The order parameters vary periodically 
and in phase along the radial direction 
during the growth of the spherulite. 
This in phase variation of the order parameters
arises due to their bilinear coupling with $\gamma >0$ in 
equation~\ref{eq.fed}.

In these numerical computations, a
banded spherulite grows from  an initial nano crystallite seed on 
sufficient supercooling of the smectic phase. 
The nano crystallites have higher density compared to the amorphous state. 
Thus the growth of this nano 
crystallite-rich domain leads to decrease in density around it due to the 
depletion of the molecules. When the  
density decreases sufficiently, it promotes the formation 
of a crystallite-poor amorphous domain around the initial crystallite-rich 
domain. 
The growth of this crystallite-poor band in 
turn increases the density around its periphery and leads to the nucleation 
of another nano crystallite-rich band. 
The growth continues with the formation of alternating crystalilte-rich and 
crystalilte-poor bands in a periodic manner 
along the radial direction of the spherulite as indicated in 
Fig.~\ref{fig:FDSimEg0}.
The radius of the numerically simulated banded spherulite grows with time
in a rhythmic fashion as shown in the Supplementary Material Fig.~S7.

The spacing between two successive bands of the spherulite was also 
calculated from the numerical
results for different values of supercooling $\Delta$T. The numerically 
computed band spacing as a function of 
supercooling agrees very well with the experimental data as shown in 
Fig.~\ref{fig:Sigma}c.
This figure also shows that the experimentally measured band spacing
of the spherulite tends to diverge on 
approaching $\Delta$T $\sim$ 22.5 K from above.
Below this supercooling, only the formation of non-banded spherulites was 
observed experimentally.   
Both the LSA and numerical results 
account for this divergence of band spacing. 
This divergence arises from a singularity of the TDGL equations for the 
parameter $\epsilon$ 
in equations~\ref{TDGLphieqn} and \ref{TDGLpsieqn} being zero.
Therefore, the order parameter $\psi$ is determined by the equation
$\psi(\psi-1)(\psi-\alpha)=\frac{\gamma}{W}\phi$ and only the order 
parameter $\phi$ controls the dynamics of the system. A comparison  between 
the computed and experimentally
measured growth velocities of banded spherulite gives the translational 
diffusion constant ($A\Gamma_\phi$) of 
the molecules as 10$^{-11}$ m$^2$/sec which agrees with the earlier 
reported data on 8OCB \cite{dvinskikh2012}. 
The consistency of the numerical results was also checked on solving the 
TDGL equations using the finite element 
method (FEniCS python package)(see  Fig.~S8 in the Supplementary Material). 
However, the numerical solutions of these 
nonlinear equations show that the ring 
banded pattern under certain conditions becomes unstable leading to the 
breaking of the spherical symmetry of the 
growth pattern. The detail on these numerical results will be published 
elsewhere.

\begin{widetext}
\begin{minipage}{\linewidth}
\begin{figure}[H]
\centering
\includegraphics[width=17.0 cm]{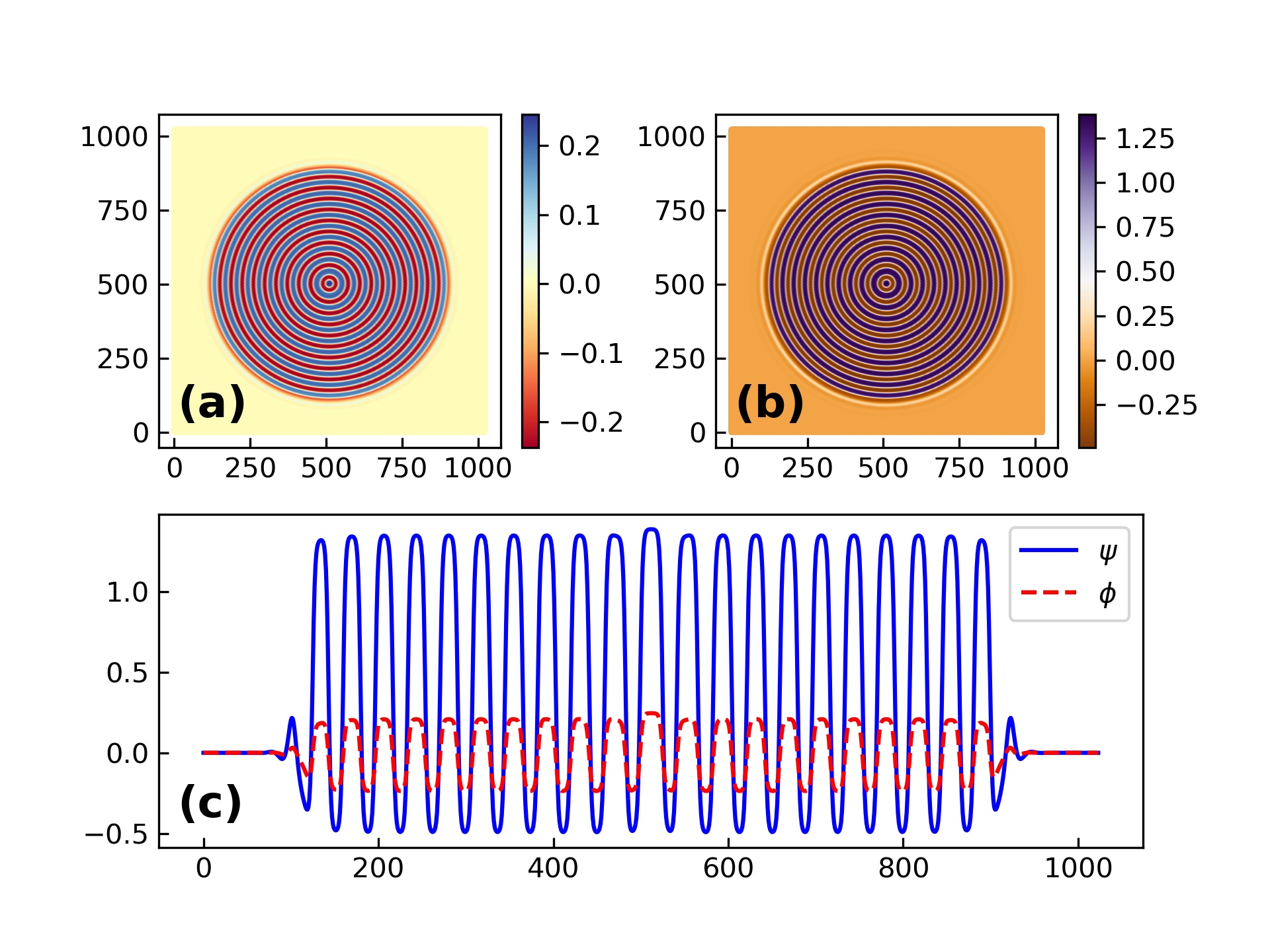}
 \caption{\textbf{Simulated banded spherulite domain.} 
 The colour coded representation of the order parameters (a) $
 \phi$  and (b) $\psi$  of a banded spherulite 
 obtained from the numerical solution of the TDGL equations at $\Delta$T = 
 39.5 K. The model parameters
 used are $\gamma /A = 0.242$, $W/A = 0.1$, $\zeta_\phi=0.04$ $\mu$m, $
 \zeta_\psi=3.16\zeta_\phi$, 
 $\lambda_D=1.41\zeta_\phi$, time step size $dt = 0.01$ and spatial step 
 size $dx=dy=1.0$. 
(c) The graphical profiles of $\phi$ and $\psi$ along a radial axis of the 
simulated banded spherulite.}
\label{fig:FDSimEg0}
\end{figure}
\end{minipage}
\end{widetext}

\section{Summary }
We have studied the banded spherulitic growth of the solid phase of a pure 
liquid crystalline compound from its melt. The compound exhibits 
coexistence of fibrillar nano crystallites and an amorphous phase in its 
most stable solid state at room temperature. 
The banded spherulites are formed due to the rhythmic generation of 
crystallite-rich and crystallite-poor amorphous
concentric zones on supercooling the smectic phase of the compound. This 
produces
the modulation in birefringence which gives rise to the interference colour 
bands observed between crossed
polarisers. A theoretical model using the time dependent Ginzburg-Landau 
theory is developed to account for the observed
banded spherulitic growth in this compound which agrees well with the 
experimental observations. 

\begin{acknowledgments}
We thank Ms Vashudha K. N. for her help in acquiring
XRD data and K. M. Yatheendran for his help in FESEM and confocal imaging.
\end{acknowledgments}

\end{document}


\maketitle
\renewcommand{\thefigure}{S\arabic{figure}}
\begin{center}
\Large{ \textbf{Supplementary Material }}   
\end{center}

\section*{Sample preparation for Cryo-SEM imaging}
  For cryo-SEM imaging, a small brass container was filled with the sample above its clearing temperature using  
a hot stage.  After that it was dipped in liquid nitrogen for fast cooling which produced the banded 
spherulites. It was then transferred to the cryo-sample preparation chamber (Quorum Technologies PP3000T) of the 
FESEM system maintained at a temperature of 103 K. The sample was then heated to 183 K to make it soft and 
the upper portion of the sample was scraped off for removing the ice content.
The newly opened sample surface was then sputter coated with platinum and transferred to the main unit for 
SEM imaging where the sample was maintained at 83 K. 

\section*{Theoretical Discussions}
We provide here the details of the linear stability analysis and numerical methods for the solution of the time dependent 
Ginzburg-Landau model C equations of the banded spherulitic growth in our sample. In the main text,  
two order parameters are defined to describe the banded
spherulitic growth. The conserved order parameter $\phi = [\rho - \rho_0 ]/\rho_0$ describes the local deviation of density 
from the average density $\rho_0$ in the smectic phase and the non-conserved order parameter 
$\psi = (\rho_ {nc} - \rho_a )/\rho_0$ describes the local composition. The parameters $\rho_{ nc} /\rho_0$ and 
$\rho_a /\rho_0$ are the fractional density of molecules in the nano crystallite and the amorphous solid phase respectively. 
The free energy density of the system in terms of these order parameters  can be written as
\begin{equation}
f=\frac{A}{2}\phi^2 +\frac{1}{2}k_\phi |\nabla \phi|^2 + W \Big[ \frac{\alpha(T)}{2}\psi^2 - \frac{1+\alpha(T)}{3}\psi^3 
   +\frac{1}{4}\psi^4 \Big]+\frac{1}{2}k_\psi |\nabla \psi|^2-\gamma \phi \psi.
\label{eq.fd}
\end{equation}
The temperature dependent parameter $\alpha$ assumed as $0\leq\alpha(T)\leq 1$ drives to the first order transition in the
system.  For simplicity, all other coefficients of the expansion  are assumed to be temperature independent constants in 
our model.  The time dependent Ginzburg-Landau (TDGL) Model C equations for the order parameters are given by
\begin{equation*}
\frac{\partial \phi}{\partial t}=\Gamma_\phi \nabla^2 \frac{\delta F}{\delta \phi},
\end{equation*}

\begin{equation*}
\frac{\partial \psi}{\partial t}=-\Gamma_\psi  \frac{\delta F}{\delta \psi},
\end{equation*}
where right hand side of these equations is obtained by taking the variational derivative of 
the free energy F with respect to the order parameters.

Using the free energy density in eq.~\ref{eq.fd}, the dynamical equations for the order parameters become
\begin{equation*}
\frac{\partial \phi}{\partial t}=\Gamma_\phi \nabla^2 \Big[A\phi -\gamma\psi -k_\phi  \nabla^2 \phi \Big] 
\end{equation*}
\begin{equation*}
\frac{\partial \psi}{\partial t}=-\Gamma_\psi \Big[W\psi(\psi-1)(\psi-\alpha) -\gamma\phi -k_\psi  \nabla^2 \psi  \Big].
\end{equation*}
The above equations can be made dimensionless by the transformations  
$x\rightarrow x^\prime=\frac{\sqrt{\epsilon}}{\xi_\phi}x$, 
$t \rightarrow t^\prime=\frac{\epsilon \Gamma_\phi A}{\xi^2_\phi}t$.
The dimensionless form of these equations after omitting for convenience the prime notation on the variables becomes
\begin{eqnarray}\label{TDGL_equn}
\frac{\partial \phi}{\partial t} &=& \nabla^2 \Big[\phi -\frac{\gamma}{A}\psi -\epsilon \nabla^2\phi  \Big]\\
\frac{\partial \psi}{\partial t} &=& -\frac{\xi^2_\phi}{\epsilon \lambda^2_D} \Big[\psi(\psi-1)(\psi-\alpha)-\frac{\gamma}{W}\phi -\epsilon \frac{\xi^2_\psi}{\xi^2_\phi}\nabla^2\psi  \Big],
\end{eqnarray}
where the parameters are $\epsilon=\frac{\gamma^2}{WA} -\alpha$, $\xi_\phi=\sqrt{\frac{k_\phi}{A}}$ , $\xi_\psi=\sqrt{\frac{k_\psi}{W}}$ and $\lambda^2_D = \frac{A\Gamma_\phi}{\Gamma_\psi W}$. 
As usual in Landau theory, the parameter $\alpha$ is assumed to be linearly dependent on temperature as
$\alpha=\delta (T-T_c)$. The melting temperature of 8OCB is 327.6 K and smectic phase is absolutely stable above this 
temperature. Hence, we have considered that $\alpha$ takes the maximum value 1 at this temperature $T_m=327.6$ K. 
We have assumed 
the temperature $T_c=273.1$ K denoting the lower limit of stability of the smectic phase which is difficult to find experimentally 
due to large range of supercooling of the smectic phase. Then the parameter  $\alpha(T)=(T-T_c)/(T_m -T_c)$.

\subsection*{Linear Stability Analysis}
The linear stability analysis (LSA) was performed to investigate the growth of banded spherulite on quenching the system 
from higher temperature smectic phase. The TDGL equations were linearised with respect to small perturbations 
about the fixed point ($\phi^*=0$, $\psi^*=0$). Hence, we get the following linear dynamical equations by considering 
$\phi(\vec{r},t)=\phi^* +\delta \phi(\vec{r},t)$ and $\psi(\vec{r},t)=\psi^* +\delta \psi(\vec{r},t)$ . 
\begin{eqnarray*}
\frac{\partial \delta \phi}{\partial t} &=& \nabla^2 \Big[\delta \phi -\frac{\gamma}{A} \delta\psi -\epsilon \nabla^2 \delta\phi  \Big]\\
\frac{\partial \delta \psi}{\partial t} &=& -\frac{\xi^2_\phi}{\epsilon \lambda^2_D}\Big[g(\psi^*)-\frac{\gamma}{W}\phi^*\Big] -\frac{\xi^2_\phi}{\epsilon \lambda^2_D} \Big[g^\prime(\psi^*)\delta \psi 
-\frac{\gamma}{W} \delta \phi -\epsilon \frac{\xi^2_\psi}{\xi^2_\phi}\nabla^2 \delta \psi  \Big]
\end{eqnarray*}
where $g(\psi)=\psi(\psi-1)(\psi-\alpha)$ and $g^\prime(\psi)$ is the derivative of $g(\psi)$ with respect to $\psi$. In our model $g(0)=0$ and $g^\prime (0)=\alpha > 0$.
Assuming the solution of the above two equations in the form 
$\delta\phi=\delta\phi_0 e^{\sigma t} e^{i\vec{q}\cdot \vec{r}}$ and 
$\delta\psi=\delta\psi_0 e^{\sigma t} e^{i\vec{q}\cdot \vec{r}}$,  we obtain

\begin{eqnarray*}
\sigma \delta\phi_0 e^{\sigma t} e^{i\vec{q}\cdot \vec{r}} &=& \Big[-(q^2+\epsilon q^4) \delta\phi_0 
+\frac{\gamma}{A}q^2\delta\psi_0   \Big]e^{\sigma t}e^{i\vec{q}\cdot \vec{r}} \\
\sigma \delta\psi_0 e^{\sigma t} e^{i\vec{q}\cdot \vec{r}} &=& -\frac{\xi^2_\phi}{\epsilon \lambda^2_D}
\Big[ \alpha\delta\psi_0 -\frac{\gamma}{W}\delta\phi_0 
+ \epsilon \frac{\xi^2_\psi}{\xi^2_\phi} q^2 \delta\psi_0   \Big] e^{\sigma t}e^{i\vec{q}\cdot \vec{r}}
\end{eqnarray*}
or
\begin{eqnarray*}
\sigma \delta\phi_0 &=& -(\epsilon q^4 +q^2)\delta\phi_0 +\frac{\gamma}{A} q^2 \delta\psi_0  \\
\sigma \delta\psi_0 &=&  \frac{\xi^2_\phi}{\epsilon \lambda^2_D} \frac{\gamma}{W} \delta\phi_0 -\frac{\xi^2_\phi}{\epsilon \lambda^2_D} \big[\alpha +\epsilon \frac{\xi^2_\psi}{\xi^2_\phi}  q^2 \big]\delta\psi_0
\end{eqnarray*}

These equations can be written as the eigenvalue equation 
$B \begin{pmatrix}
\delta\phi_0 \\ \delta\psi_0
\end{pmatrix}=\sigma \begin{pmatrix}
\delta\phi_0 \\ \delta\psi_0
\end{pmatrix}$,
where
\begin{equation*}
B=\begin{bmatrix}
-(q^2+ \epsilon q^4) & \frac{\gamma}{A}q^2 \\
 \frac{\gamma \xi^2_\phi}{\epsilon W \lambda^2_D}  & -\frac{\xi^2_\phi}{\epsilon \lambda^2_D} \big[\alpha +\epsilon  \frac{\xi^2_\psi}{\xi^2_\phi}  q^2 \big]
\end{bmatrix}.
\end{equation*}
The eigenvalues of the  matrix $B$ can be found from the equation $\sigma^2 -\tau \sigma + \Delta=0$, 
where $\tau=Tr(B) $ and $\Delta =det(B)$ are given by
\begin{eqnarray*}
\tau &=& -\Big[(q^2+ \epsilon q^4)  +\frac{\xi^2_\phi}{\epsilon \lambda^2_D} \big[\alpha 
+\epsilon  \frac{\xi^2_\psi}{\xi^2_\phi}  q^2\big] \Big] \\
\Delta &=& \frac{\xi^2_\phi}{\epsilon \lambda^2_D}(q^2+ \epsilon q^4) \big[\alpha 
+\epsilon  \frac{\xi^2_\psi}{\xi^2_\phi}  q^2\big] - \frac{\gamma^2 \xi^2_\phi}{\epsilon W A \lambda^2_D} q^2.
\end{eqnarray*}

The eigenvalues $\sigma_{\pm} = \frac{\tau \pm \sqrt{\tau^2 -4\Delta}}{2}$ can be complex in general.
If the real part  of any of these eigenvalues is positive, the corresponding mode grows with time ($\sim e^{\sigma t}$) and 
the fixed point is unstable. On the other hand, the fixed point is stable if the real part of both the eigenvalues is negative.
The real part of 
$\sigma_{-}$ is negative for all values of $q$ as $\tau<0$. But the real part of $\sigma_{+}$ can be positive if $\Delta <0$ giving
rise to the unstable modes.
The $\Delta$  becomes negative in the range $0<q<q_{max}$ where the limits of the range can be obtained from 
the equation $\Delta(q)=0$ or 
  \begin{equation*}
  \frac{ \xi^2_\phi }{\lambda^2_D }\bigg[\epsilon \frac{\xi_\psi^2}{\xi_\phi^2}q^6 +\Big(\alpha + \frac{\xi_\psi^2}{\xi_\phi^2}\Big)q^4 -q^2\bigg]=0
 \end{equation*}
giving the lower root $q=0$ and the upper real root
\begin{equation*}
q_{max} =\frac{1}{\sqrt{2\epsilon}}\bigg[ \sqrt{\frac{4\epsilon\xi^2_\phi}{\xi^2_\psi} +\Big(1+\frac{\alpha\xi^2_\phi}{\xi^2_\psi}\Big)^2} -\Big(1+\frac{\alpha\xi^2_\phi}{\xi^2_\psi}\Big) \bigg]^{1/2}.
\end{equation*}

The fastest growing mode  $q=q_{c}$ can be found by maximising $\sigma_+$ with respect to $q$.  The wavelength of 
this dominant growing mode
determined numerically is compared with the experimentally observed band spacing of the system for different values
of supercooling. The theoretically calculated band spacing diverges on
approaching the supercooling $\Delta T=22.5$ K 
as observed experimentally in our system.  This divergence can be attributed to the singularity of the TDGL equations at 
$\epsilon=0$ in our model. 

\subsection*{Simulation}
\begin{figure*}
   \includegraphics[width=17.8cm]{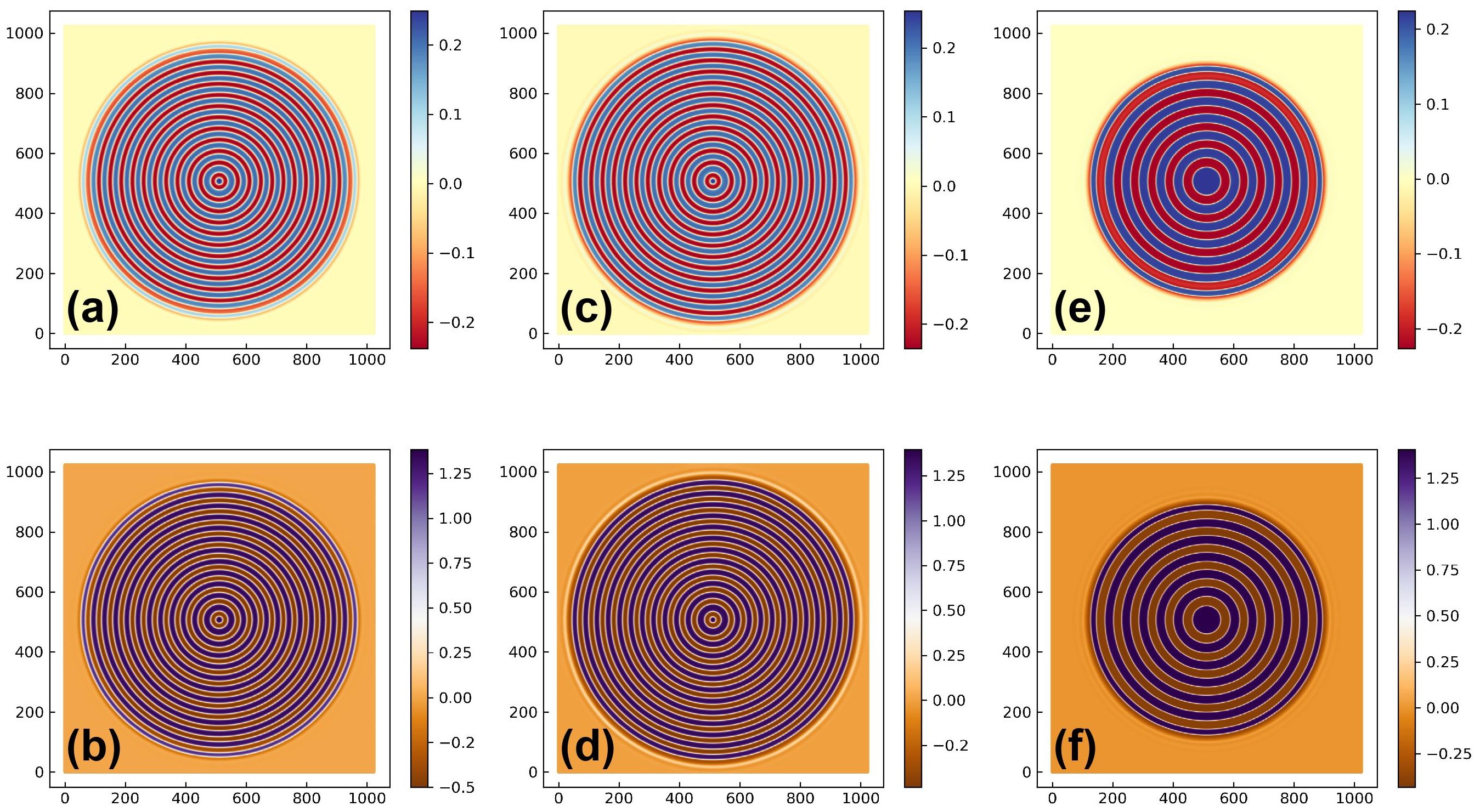}
   \centering
   \caption{\textbf{Simulated banded spherulite patterns.}
   Numerical solutions of $\phi$ (upper row) and $\psi$ (lower 
   row) 
   for different values of $\alpha$ (or $\Delta$T).  (a, b)
   $\alpha=0.250$ ($\Delta$T = 40.8 K) at $149999$ time steps, (c, d) $\alpha=0.300$ ($\Delta$T = 38.1 K) at $199999$ 
   time steps, (e, f) $\alpha=0.450$ ($\Delta$T = 29.9 K) at $399999$ time steps on a 
   $1024\times1024$ square lattice.}
     \label{fig:PhiPsi}
\end{figure*}

For numerical solutions,  we constructed  three lower order partial differential equations from the 
higher order the TDGL equations by considering an additional dynamical variable $\mu$ as
\begin{eqnarray}\label{fd_Sim_eqn}
\frac{\partial \phi}{\partial t} &=& \nabla^2 \mu  \\
\mu &=&\phi -\frac{\gamma}{A}\psi -\epsilon \nabla^2\phi \\
\frac{\partial \psi}{\partial t} &=& -\frac{\xi^2_\phi}{\epsilon \lambda^2_D} 
\Big[\psi(\psi-1)(\psi-\alpha)-\frac{\gamma}{W}\phi -\epsilon \frac{\xi^2_\psi}{\xi^2_\phi}\nabla^2\psi  \Big]
\end{eqnarray}
These equations were solved numerically for two dimensional system under the no flux boundary conditions i.e., 
$\vec{\nabla}\phi \cdot \hat{n}|_{\partial\Omega}=0$, $\vec{\nabla}\mu \cdot \hat{n}|_{\partial\Omega}=0$ and 
$\vec{\nabla}\psi \cdot \hat{n}|_{\partial\Omega}=0$, where ${\partial\Omega}$ denotes the boundary. Initially $\phi$ was 
set to zero at all points in the simulation domain and $\psi$ was set to a nonzero value for a small seed at the centre of the 
simulation domain 
which was  assumed to be a square lattice. We used the finite difference method for both space and time for the discretization of
these equations. Five points were used for discretization of the Laplacian operator. We performed the finite difference 
simulation for the parameter set ( $ W/A =0.1$, $\gamma/A=0.2423$, $\xi_\phi=0.04$ $\mu$m, 
$\xi_\psi=3.16\xi_\phi$, $\xi^2_\phi / \lambda^2_D =5W/A$) obtained from the linear stability analysis.
Some numerically obtained ring banded patterns for different values of $\alpha$ are shown in Figure~\ref{fig:PhiPsi}. 
The band 
spacing in the pattern was determined using ImageJ tools. We have compared the numerically obtained band spacing data 
with the experimental as well as the linear stability analysis data in the main article.

\begin{figure*}
   \includegraphics[width=17.8cm]{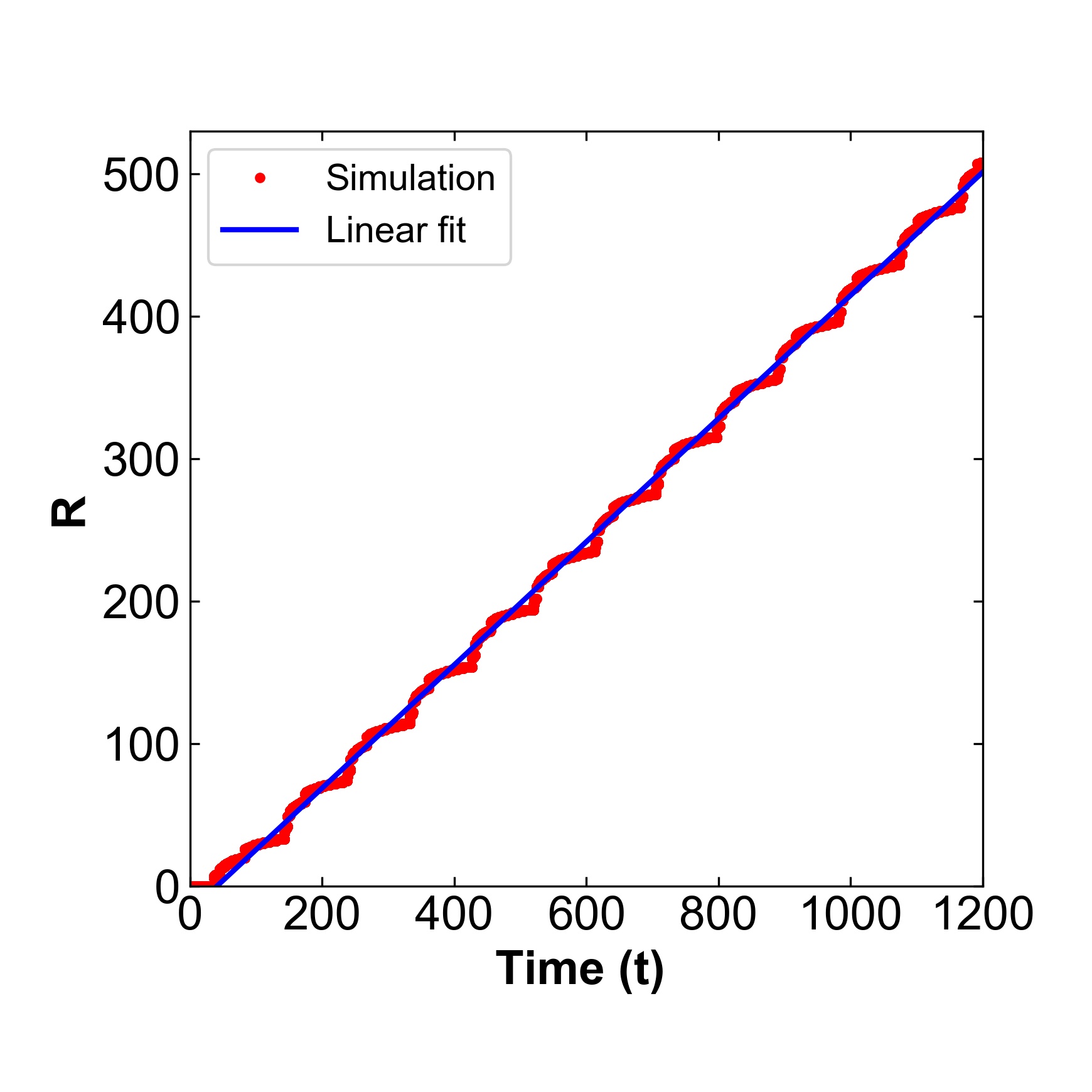}
   \centering
   \caption{\textbf{Rhythmic variation of spherulite radius.}
   Numerically computed variation of radius $R$ of the banded 
   spherulite with time showing its rhythmic growth for $\alpha=0.15$
   ($\Delta$T = 46.3 K). 
   The blue line shows the average linear growth of the spherulite with 
   time. $R$ and $t$ are dimensionless quantities. }
    \label{fig:Rvst}
\end{figure*}

The radius of the numerically simulated banded spherulite grows with time in a rhythmic fashion. The staircase type growth
of dimensionless radius $R$ with dimensionless time is shown 
in (Figure~\ref{fig:Rvst}). Each step represents the duration of a full band growth. Though this growth is
nonlinear at each individual step, it shows on average a linear growth law with time. The growth velocity of the ring 
banded pattern can be found from this linear relationship. It was not possible to resolve experimentally the rhythmic
growth pattern due to relatively fast growth of the banded spherulite in our sample. 

\begin{figure*}
   \includegraphics[width=17.8cm]{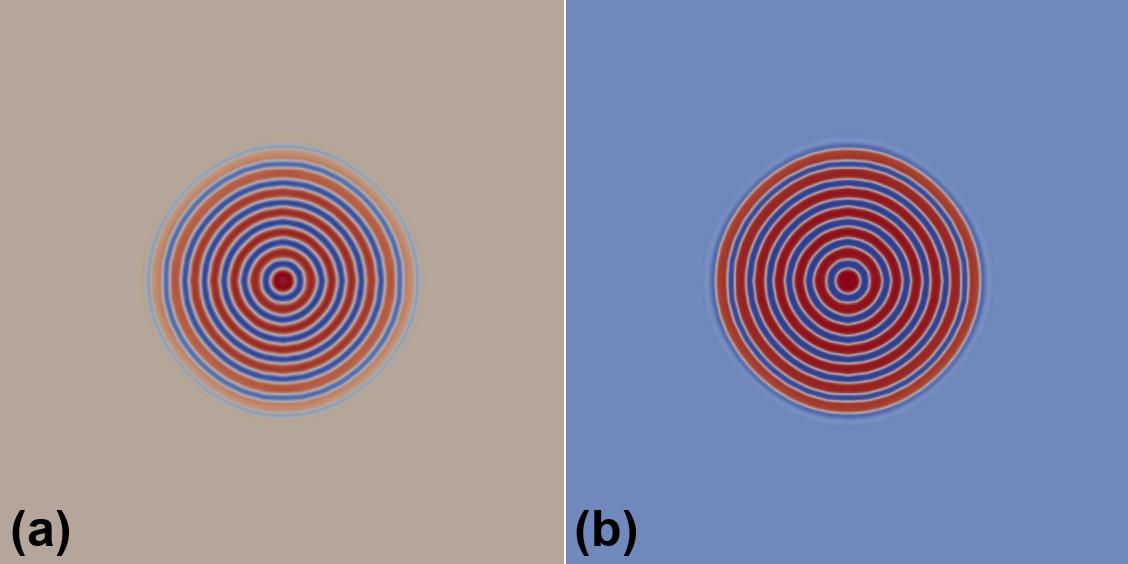}
   \centering
   \caption{
   \textbf{Banded spherulite patterns by FEM.}
   The numerical solutions of (a) $\phi$ and (b) $\psi$ 
   obtained from FEM at $2000$ time steps. 
   The parameter values used for the calculations are $\alpha=0.2$,  $
   \gamma/A=0.95$, $\xi_\psi/\xi_\phi=0.5587$, 
   $\xi^2_\phi / \lambda^2_D =20.5$ and $W/A=2.05$.  }
   \label{fig:FEM}
\end{figure*}

The TDGL equations were also solved numerically using finite element method (FEM) for some parameter values.  
We implemented 
the above equations (eq.~\ref{fd_Sim_eqn}) in the finite element open source software  
FEniCS and solved these equations on a $257\times257$ square mesh.
The FEM computations show ring banded patterns (Figure~\ref{fig:FEM}) 
similar to finite difference results 
(Figure~\ref{fig:PhiPsi}).
The FEM computations are found to be slow compared to the finite difference 
method and all the results reported in this
paper were performed in the finite difference method.


\begin{figure*}
    \includegraphics[width=17.8cm]{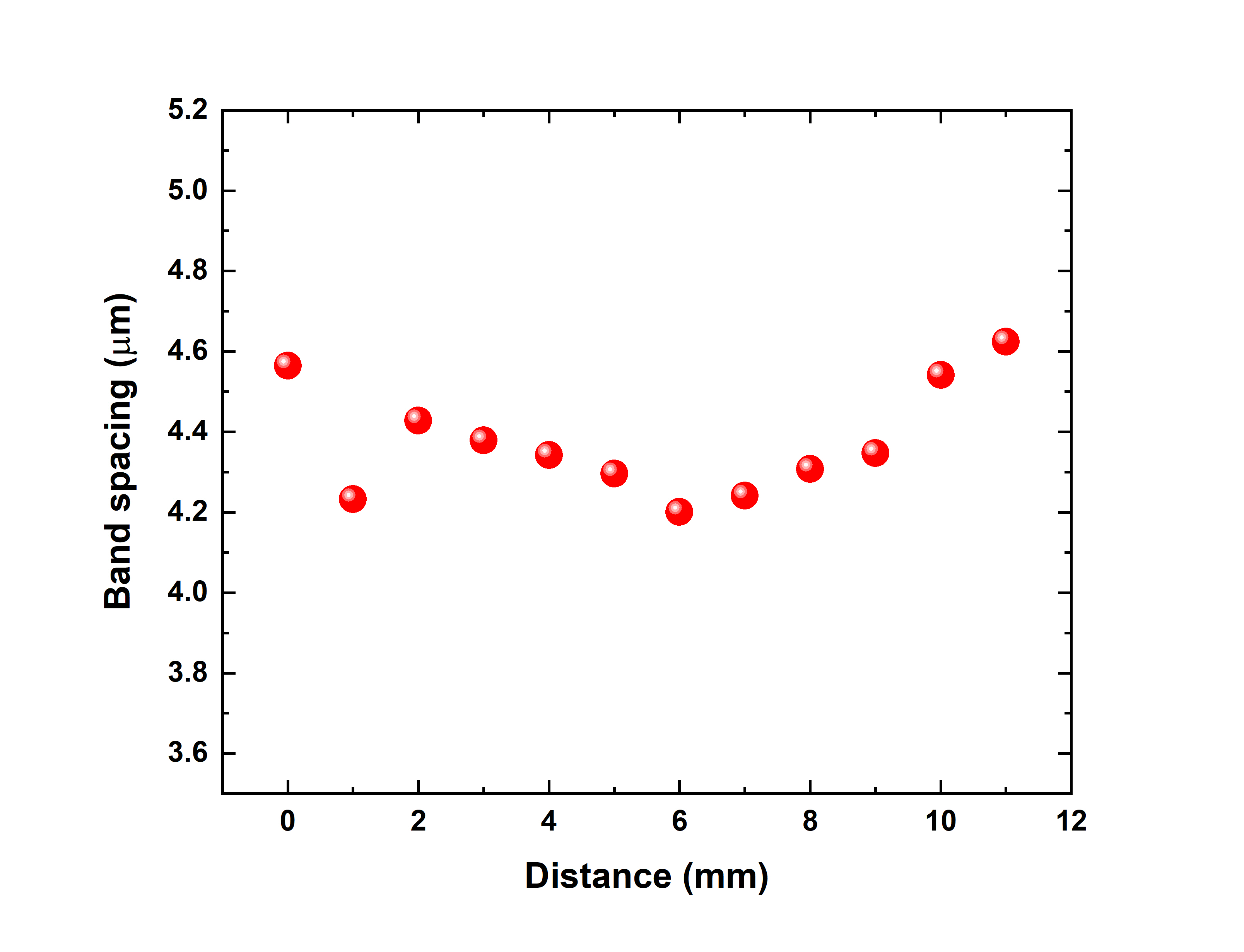}
    \centering
    \caption{\textbf{Band spacing with sample thickness.}
    The variation of band spacing of a banded spherulite of 8OCB  
    along the length of a wedge type cell with 
    increasing sample thickness. The thickness of the sample increases by 
    six times over 11 mm length of the  wedge 
    cell but the band spacing of the spherulite remains same within 
    experimental error. Thus band spacing is not
    determined by the thickness of the sample as also found in our 
    theoretical model.}
    \label{figS1} 
\end{figure*}

\begin{figure*}
    \includegraphics[width=17.8cm]{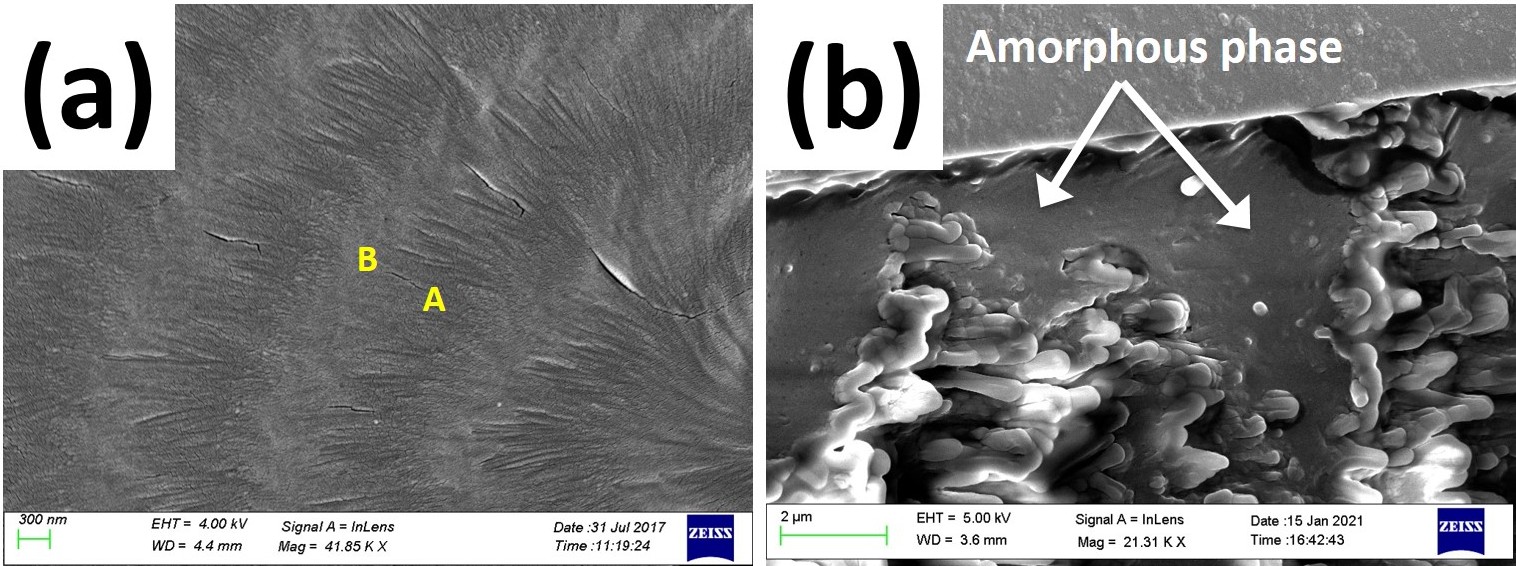}
    \centering
    \caption{\textbf{Two phase coexistence in banded spherulite.} 
    (a) The cryo-SEM image of 8OCB banded spherulite formed after 
    quenching the sample from its melt in 
    liquid nitrogen. Crystallite-rich and crystallite-poor zones are marked 
    as \textbf{A} and \textbf{B} respectively.
    (b) The SEM texture of the cross-sectional plane of the banded 
    spherulite obtained by cutting it into two halves.
    The spherulite was formed on a cleaned coverslip on cooling the sample 
    to room temperature. The amorphous domains
    in addition to the nano crystallites are clearly visible.}
    \label{figS2}
\end{figure*}

\begin{figure*}
    \includegraphics[width=17.8cm]{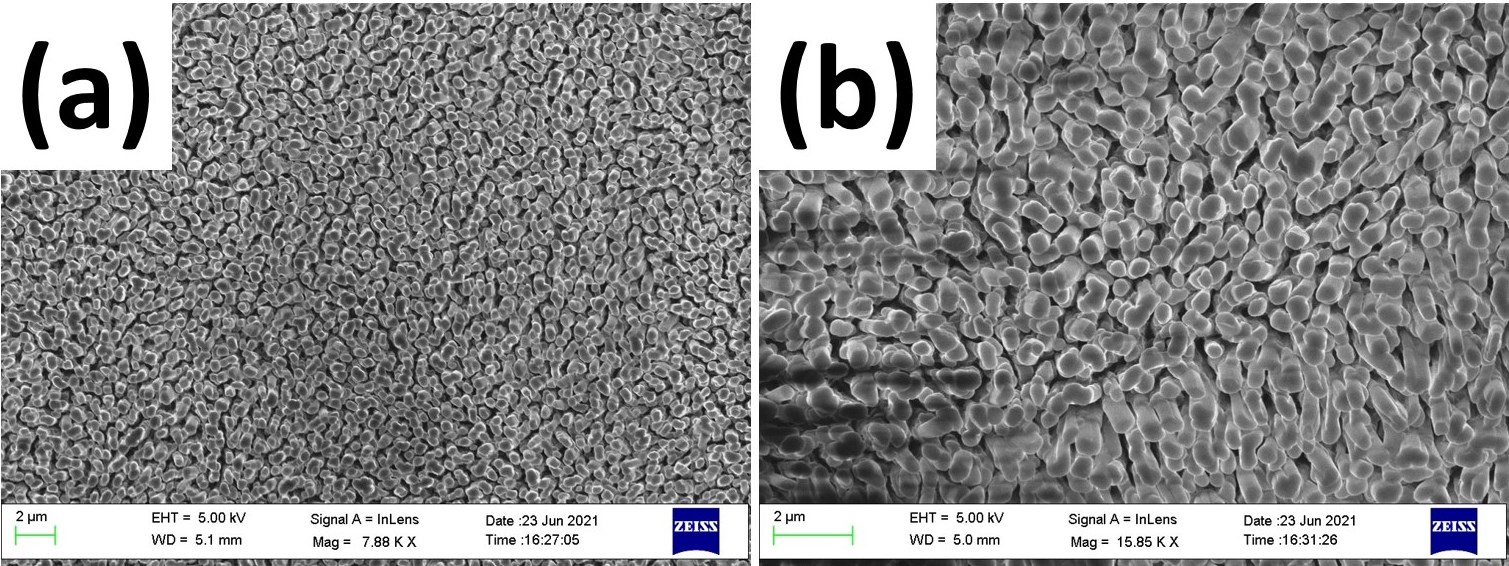}
    \centering
    \caption{\textbf{Cross sectional image of nano crystallites.} (a) The SEM
    texture of the cleaved surface of the banded 
    spherulite showing the cross sectional view
    of the fibrillar nano crystallites of 8OCB. The spherulite was formed 
    on a coverslip by cooling the sample from its
    melt to room temperature and then cut in the middle for the cross 
    sectional view of the banded spherulite domain. 
    (b) The same texture with higher magnification.}
    \label{figS3}
\end{figure*} 

\begin{figure*}
    \includegraphics[width=17.8cm]{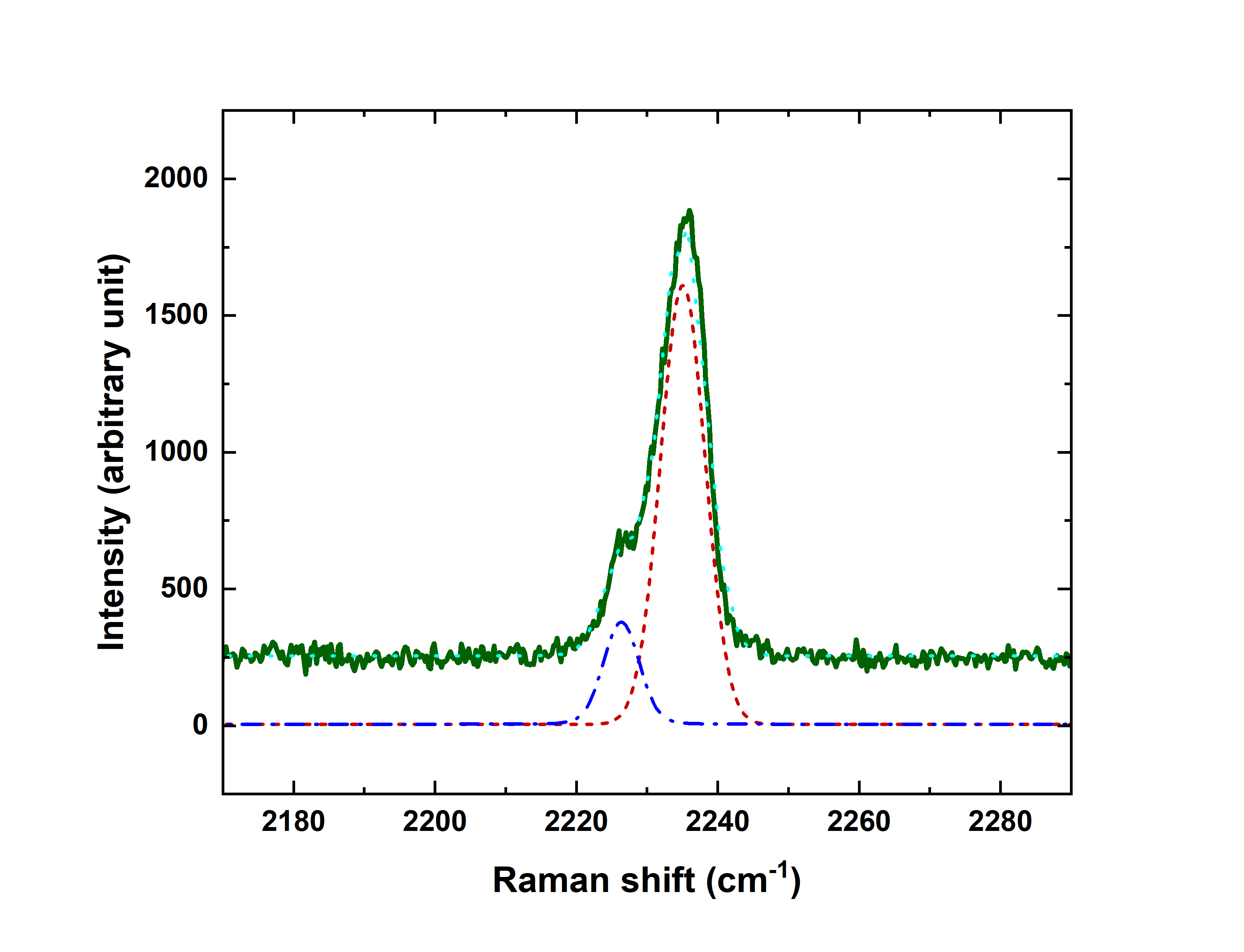}
    \centering
    \caption{\textbf{CN peak from banded spherulite.}
    The asymmetric Raman peak profile (green line) for the CN 
    stretching vibration of 8OCB in the banded 
    spherulite domain. This peak profile can be fitted with two closely 
    positioned peaks at 2226 cm$^{-1}$ (blue dash dot line) 
    and 2235 cm$^{-1}$ (red dash line) respectively. These two peaks for CN 
    stretching vibration arise from the amorphous 
    and nano crystallite components of the banded spherulite respectively.}
    \label{figS4} 
\end{figure*}

\begin{figure*}
    \includegraphics[width=17.8cm]{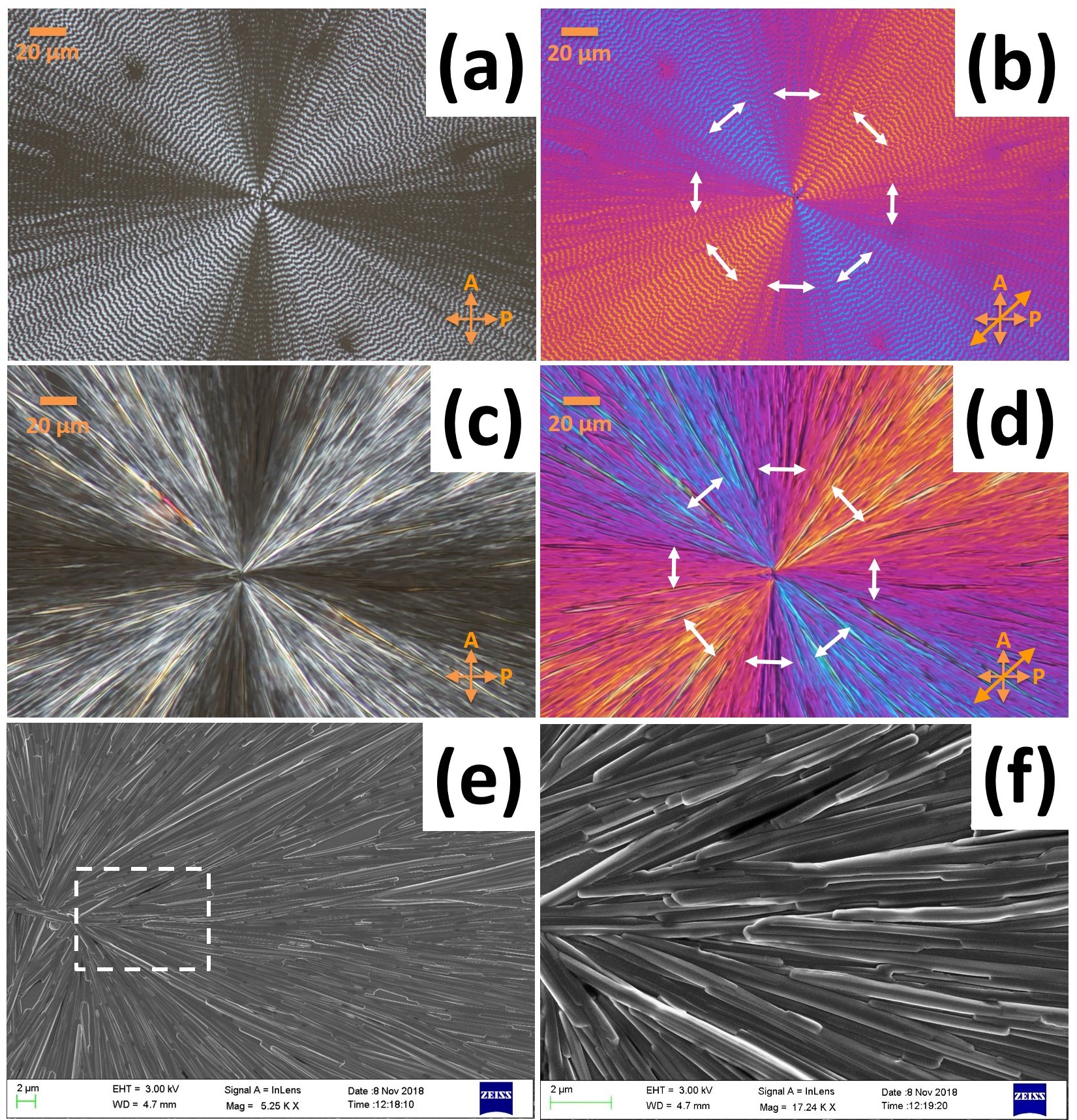}
    \centering
    \caption{ \textbf{Transformation of banded spherulite on heating.}
    The polarised optical microscope (POM) images of a banded 
    spherulite of 8OCB at room temperature
    (a) between crossed 
    polarisers and (b) with a $\lambda$-plate inserted in the optical path 
    in addition to the crossed polarisers. 
    The interference colours indicate that the spherulite is optically 
    negative. The direction of major refractive index in the 
    sample plane is denoted by white double headed arrows around the seed. 
    (c) The POM texture after holding the
    banded spherulite at 325.1 K for 1 hour followed by cooling it to room 
    temperature. 
    Banded spherulite changes irreversibly to a non-banded one. Dark arms 
    of the Maltese 
    cross parallel to the polarisers remain invariant on rotation of the 
    sample between crossed polarisers. 
    (d) The same texture as shown in (c) after inserting a $
    \lambda$-plate in the optical path. 
    The interference colours indicate that 
    the non-banded spherulite remains 
    optically negative with the
    major refractive index along the white double headed arrows around the 
    seed.    
    (e) The FESEM image of the same sample as in (c) showing 
    the radially aligned fibrillar nano crystallites 
    with no periodic variation of their density. The crystalline fibrils 
    become longer on transformation of the banded 
    spherulite to the non-banded one. 
    (f) The magnified view of the selected portion in (e) 
    depicting no twist in the nano crystalline fibrils.} 
    \label{figS5}
\end{figure*}

